\documentclass[twocolumn,prl,showpacs,preprintnumbers,amsmath,amssymb]{revtex4}
\usepackage{graphicx}
\usepackage{dcolumn}
\usepackage{bm}

\begin{document}

\preprint{APS/123-QED}
\title{Phase-sensitive detection of Bragg scattering at $1D$ optical lattices}

\author{S. Slama}
\author{C. von Cube}
\author{B. Deh}
\author{A. Ludewig}
\author{C. Zimmermann}
\author{Ph.W. Courteille}
\affiliation{Physikalisches Institut, Eberhard-Karls-Universit\"at T\"ubingen,\\
Auf der Morgenstelle 14, D-72076 T\"ubingen, Germany}

\date{\today}

\begin{abstract}
We report on the observation of Bragg scattering at $1D$ atomic lattices. Cold atoms are confined by optical dipole forces at the 
antinodes of a standing wave generated by the two counter-propagating modes of a laser-driven high-finesse ring cavity. By heterodyning 
the Bragg-scattered light with a reference beam, we obtain detailed information on phase shifts imparted by the Bragg scattering process. 
Being deep in the Lamb-Dicke regime, the scattered light is not broadened by the motion of individual atoms. In contrast, we have 
detected signatures of global translatory motion of the atomic grating.
\end{abstract}

\pacs{42.50.Vk, 05.45.Xt, 05.65+b, 05.70.Fh}

\maketitle
	
Elastic Rayleigh scattering is a phase-coherent process. In thermal atomic clouds however the photonic recoil transferred to the atoms 
by the scattering process introduces inhomogeneous line broadening washing out the phase-coherence. Furthermore, Rayleigh scattering 
only dominates at small intensities; stronger pumping causes power broadening, which merely increases the inelastic components of the 
resonance fluorescence. For these reasons it is difficult to directly measure the phase-shift induced by Rayleigh scattering. 

A phase-coherent study of Rayleigh scattering is facilitated by two measures: 1. Using cold and strongly confined atoms, and 2. 
arranging for long-range order in the atomic cloud. By cooling the atoms the Doppler-broadening is reduced; by tightly confining them 
within a very small region of space, i.e. within the Lamb-Dicke limit, no recoil is imparted to individual atoms; and by creating density 
gratings the elastic part of the resonance fluorescence is concentrated into a very small solid angle. The resonant enhancement of the 
structure factor for elastic Rayleigh scattering by atomic long-range order is called Bragg scattering. 

Periodic structures are generally probed by Bragg scattering. A probe beam is shone onto the sample under a certain angle, the so-called 
Bragg angle, and the emergence of phase-coherent light at well-defined sharp solid angles is a signature of long-range order. This 
procedure can be applied to periodic arrangements of atoms in optical gratings. Bragg scattering of light at $3D$ atomic lattices has 
for the first time been observed by Birkl \textit{et al.} and Weidem\"uller \textit{et al.} \cite{Birkl95,Weidemuller95}. 

The elastic peak of the atomic response to incident laser light has been observed in several experiments. Westbrook \textit{et al.} 
\cite{Westbrook90,Jessen92} used the heterodyne method to beat down the fluorescence of magneto-optically trapped atoms with a local 
oscillator to electronically accessible frequencies. It is in principle possible to probe the \textit{complete} fluorescence spectrum, 
i.e. the Mollow triplet and the elastic Rayleigh peak by scanning the reference laser. This technique permitted Westbrook 
\textit{et al.} to detect Dicke-narrowing of Rayleigh scattering in magneto-optical traps. However in this experiment the heterodyne 
signal was integrated over long times, so that the phase-coherence of the elastic scattering process is not directly shown. 

We report here the first detailed phase information of the Rayleigh scattering process. The progress was possible by combining the 
techniques of heterodyning and Bragg scattering. In contrast to previous experiments, we exploit the heterodyne technique 
interferometrically: The frequency beat between the Bragg reflected light and a reference laser having a different frequency is 
demodulated. From the two quadrature phases, we completely recover the complex scattering coefficient. We ramp the frequencies of both 
laser beams, the Bragg beam and the reference beam simultaneously, so that the beat frequency is fixed. Because the elastically scattered 
light is always phase-coherent, we can look for phase shifts due to the scattering process. When we ramp the laser frequencies, we 
basically probe the resonant enhancement of the Rayleigh peak close to the Bragg condition. At the same time our experiment represents 
the first observation of Bragg scattering at $1D$ atomic density gratings. With a Lamb-Dicke factor on the order of $100$ we are deep 
in the Lamb-Dicke regime. 

\begin{figure}[ht]
		\centerline{\scalebox{0.35}{\includegraphics{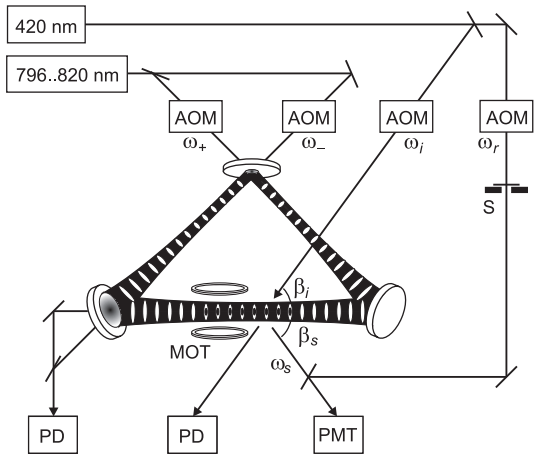}}}\caption{
		The experimental setup consists of a ring cavity pumped at $796$ - $820~$nm and a diode laser at $420~$nm for Bragg scattering. 
		The shutter $S$ controls the reference beam used to detect frequency beats between the Bragg and the reference beam.}
		\label{Fig1}
		\end{figure}

The optical layout of our experiment is shown in Fig.~\ref{Fig1}. It consists of a high-finesse ring cavity, which has been discussed 
in Ref.~\cite{Kruse03} and a setup for Bragg scattering. The ring cavity has a finesse of $80000$ and a waist of $w_{dip}=130~\mu$m. 
From a titanium-sapphire laser operating at $\lambda_{dip}=796$ - $820~$nm two light frequencies $\omega_{\pm}$ are generated by means 
of acousto-optic modulators (AOM). The light beams pump the two counter-propagating modes of the ring cavity near resonance, thus 
forming a standing wave, which propagates at a velocity $v$ given by $2k_{dip}v=\omega_+-\omega_-$. The intracavity power is 
$P_{cav}=1$ - $10~$W. Typically $N_{tot}=10^7$ $^{85}$Rb atoms are loaded from a standard magneto-optical (MOT) trap into the standing 
wave, which is red-detuned with respect to the rubidium $D_1$ line. Typically the temperature of the cloud is on the order of few 
$100~\mu$K.

The light of a blue laser diode (Toptica LD-0405-0005-2) operating at $\lambda_{brg}=420.2~$nm is split into a probe beam, $\omega_i$, 
and a reference beam $\omega_r$. The frequencies of the beams are controlled by means of AOMs. Some time after loading the atoms into 
the standing wave, the light beam $\omega_i$ is pulsed and shone under an angle of $\beta_i=58^\circ$ onto the atoms. The light 
reflected from the atoms, $\omega_s$, is detected under the angle $\beta_s=-\beta_i$ with a photomultiplier (PMT) (Hamamatsu 1P28) 
terminated with a resistive load of $R=100~$kHz. Some experiments where performed by carefully phase-matching the Bragg beam with a 
reference beam $\omega_r$ (then the shutter $S$ is open). In this case, the PMT signal was amplified with a transimpedance amplifier 
(FEMTO DHPCA-200). 

Our standing wave dipole trap represents a $1D$ optical lattice, whose periodicity is $\frac{1}{2}\lambda_{dip}=\pi/k_{dip}$. The Bragg 
condition requires $\lambda_{dip}\cos\beta_i=\lambda_{brg}$. To resonantly enhance the Bragg scattering, which otherwise would be 
neglegibly small, we tune the laser to the transition $5S_{1/2},F=3\rightarrow6P_{3/2},F'=2,3,4$ with a natural linewidth of 
$\Gamma_{brg}/2\pi=1.3~$MHz \cite{Weidemuller98}. During the Bragg pulse sequence the repumping laser of the magneto-optical trap is 
shone onto the atoms to minimize optical pumping into the ground state $F=2$ level. 

The efficiency of Bragg scattering depends critically on the angle of incidence $\beta_i$: The acceptance angle is about $0.1^\circ$. The 
scattered light beam has a nearly Gaussian elliptical shape. It is collimated in the scattering plane having about the same diameter as 
the input beam $w_z=250~\mu$m. This means that about $N_s\simeq630$ planes of the atomic lattice are illuminated, the lattice itself 
being longer. Therefore only a small fraction, $N\simeq N_{tot}/16$, of the atoms confined in the dipole trap are illuminated by the 
Bragg beam. The radial size of the atomic cloud, $w_r\approx 30~\mu$m, determines the scattered beam divergence in the direction 
orthogonal to the scattering plane. We calculate the solid angle $d\Omega_s$ in the far field, where both beams are divergent and the 
grating can be considered a point source as $\Omega_s\equiv2\lambda_{brg}^2/\pi w_r w_z\approx 1.5\times 10^{-5}$.

\begin{figure}[ht]
		\centerline{\scalebox{0.45}{\includegraphics{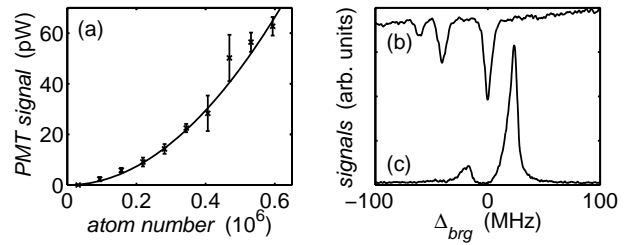}}}\caption{
		\textbf{(a)} The quadratic dependence of the Bragg scattered power $P_s$ on the atom number $N$ agrees well with 
			Eq.~(\ref{EqBragg}).
		\textbf{(b)} Absorption spectrum of magneto-optically trapped atoms showing the hyperfine levels $F'=2,3,4$. 
		\textbf{(c)} Bragg reflection spectrum obtained by ramping the blue laser frequency. Here the shutter $S$ in 
			Fig.~\ref{Fig1} blocked the reference beam. Only the two strongest hyperfine peaks are visible.}
		\label{Fig2}
		\end{figure}

The power $P_s$ diffracted by Bragg scattering into a direction $d\Omega_s$ can be estimated from \cite{Weidemuller98}:
	\begin{equation}
	\frac{dP_s}{d\Omega_s}=I_i~\frac{\pi^2}{\lambda_{brg}^4}\left|\frac{\alpha}{\epsilon_0}\right|^2\sin^2\xi~
		\left|\sum\nolimits_me^{i\Delta\mathbf{kR}_m}\right|^2~f_{DW}^2~,
	\label{EqBragg}
	\end{equation}
where 
	\begin{equation}
	\alpha=\frac{6\pi\epsilon_0}{k_{brg}^3}~\frac{\Gamma_{brg}}{2\Delta_{brg}+i\Gamma_{brg}}
	\label{EqPolarizability}
	\end{equation}
is the frequency-dependent complex polarizability. $\Delta_{brg}$ is the Bragg laser detuning. We actually use $I_i=1~$mW/cm$^2$ 
incident intensity, which is about half the saturation intensity $I_s$ and corresponds to the power $P_i=\frac{\pi}{2}w_z^2 I_i
\approx1~\mu$W. The sum over individual atoms represents the structure factor, $\sum\nolimits_m e^{i\Delta \mathbf{kR}_m}=
\sum\nolimits_m e^{i2mk_{brg}\lambda_{dip}\cos\beta}=N_s$. The Bragg scattered light power is proportional to the square of the atom 
number as is verified in Fig.~\ref{Fig2}(a). 

The Debye-Waller factor is given by $f_{DW}=\overline{e^{i\Delta\mathbf{kx}}}=e^{-\frac{1}{2}(\Delta k_z)^2\bar{z}^2}$, since only the 
distribution of atoms along the lattice normal axis $\hat{z}$ contributes to the Debye-Waller factor, $\Delta k_{x,y}=0$ and 
$\Delta k_z=2k_{brg}\cos{\beta_i}$. The \textit{rms}-size of the atomic cloud is $\bar{z}=k_{dip}^{-1}\sqrt{k_BT/2U_0}$ in the 
harmonic approximation of the trapping potential. We noticed in earlier experiments \cite{Kruse03,Nagorny03} that the temperature 
of the cloud tends to adopt a fixed ratio with the depth of the dipole trap, $T\approx0.2~U_0$. Therefore, the spatial distribution 
of the atoms (and thus the Debye-Waller factor) does not vary much with temperature, so that we estimate $f_{DW}=0.8$. Finally, 
$\xi$ is the angle between the polarization of the incident light and the diffracted wavevector. In our case, since the incident light 
is $p$-polarised with respect to the atomic grating, $\xi=90^{\circ}$. With the above estimations we calculate for the scattered power 
on resonance from Eq.~(\ref{EqBragg}) $P_s\approx 400~$nW. 

Fig.~\ref{Fig2}(c) shows a spectrum of the Bragg resonance obtained by ramping the laser frequency at $420~$nm 
(with the shutter $S$ closed and $\omega_+=\omega_-$). Three hyperfine components are expected in the spectrum, i.e. $F'=2,3,4$. 
However, the $F'=2$ component is to weak to be seen. The other lines have a relative line strength of $1:3$ and are separated by 
$40~$MHz. Fig.~\ref{Fig2}(b) shows a MOT absorption spectrum for reference. The frequency displacement between the 
spectra is due to the light shift of the atoms in the dipole trap. 

The measured peak intensity of the Bragg reflected light is on the order of $\tilde{P}_s=100~$pW. During a scan the laser is swept 
over the resonance within a time of $\Delta t=1~$ms, which is sufficient to scatter about $\Delta tP_s/2\hbar\omega_s\approx200~000$ 
photons. Defining the reflectivity as the ratio of the scattered power and the fraction of power incident on the atoms (i.e. reduced 
in order to account for the partial overlap between the incident beam and the atomic cloud), $R=\tilde{P}_s/\left(\frac{1}{2}
\pi w_rw_zI_i\right)$, we obtain for the amplitude reflection coefficient, $|r|=\sqrt R\approx 3~\%$.

The discrepancy between the calculated and the measured Bragg-reflected power, which has also been observed in \cite{Weidemuller95}, 
is due to a combination of two effects: First, the dipole-trapped atoms are subject to a position-dependent Stark shift, which 
inhomogeneously broadens the Bragg spectrum. This effect, which depends on the potential depth $U_0$ and the temperature $T$, leads to 
asymmetric resonance peaks, as seen in Fig.~\ref{Fig2}(c). From calculations we estimate a line broadening of about $10\Gamma_{brg}$ 
resulting in a strong reduction of the Bragg-reflected light peak intensity. Second, incoherent processes occur at a rate of 
$(I_i/I_s)~(1+4\Delta_{brg}^2/\Gamma_{brg}^2)^{-1}\lesssim0.5$ times the elastic scattering rate. These processes cause heating and 
optical hyperfine pumping. In fact, we observe noticeable depletion of the lattice when scanning the blue laser over the resonance. 
Experimentally, we reduce the light power and increase the scanning speed to avoid distorsion of the line profile due to heating during 
a scan.

\begin{figure}[ht]
		\centerline{\scalebox{0.43}{\includegraphics{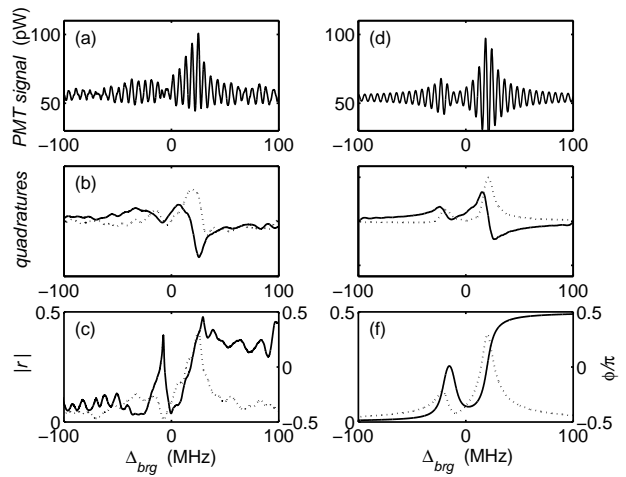}}}\caption{
		\textbf{(a)} Beat signal recorded while tuning the blue laser across resonance. The frequencies were chosen such that 
			$\Delta_i=(2\pi)5.4~$kHz. 
		\textbf{(b)} Quadrature components of the beat signal and 
		\textbf{(c)} amplitude profile calculated from the quadrature components and phase profile as obtained by counting the number of 
			oscillations in Fig.~(a) per time interval.
		\textbf{(d)} Simulated beat signal spectrum between Bragg-scattered light and a reference beam using a calculated amplitude and 
		phase profile.
		\textbf{(e)} Quadrature components of the Bragg beat demodulated with the reference beat.
		\textbf{(f)} Amplitude (dashed) and phase profile (solid) as derived from the quadrature components. The profiles coincide with 
			the profiles used to calculate the curves in Fig.~(d).}
		\label{Fig3}
		\end{figure}

The spectroscopy detailed above only yields the absolute value of the reflection coefficient $|r|$. To also measure its phase, we 
phase-match the Bragg beam with a reference laser beam and observe the frequency beat on the PMT signal. By passing the beam shone onto 
the atomic cloud, $\omega_i$, and the reference beam, $\omega_r$, through acousto-optic modulators (see Fig.~\ref{Fig1}), we can 
arbitrarily chose the beat frequency. We expect to see the frequency component $\Delta\omega_i\equiv\omega_i-\omega_r$ in the beat 
signal. A typical spectrum is shown in Fig.~\ref{Fig3}(a). 

Laser frequency fluctuations limit the resolution of the spectrum. On a long time-scale the laser emission bandwidth is estimated to 
less than $5~$MHz. However, the time-scale on which the spectrum is recorded (a few milli-seconds) is so short, that acoustic noise does 
not completely inhibit phase-sensitive detection.

When light is scattered at an unbound atomic cloud, the elastic Rayleigh peak is Doppler-broadened by the recoil imparted to the atoms, 
whose velocities have a Maxwell-Boltzmann distribution. However in axial direction the atoms are localized to less than 
$\lambda_{dip}/2\ll\lambda_{brg}$, so that the resonance fluorescence spectrum is Dicke-narrowed. The rate of inelastic scattering events 
in which the vibrational quantum number changes, is reduced by the Lamb-Dicke factor $(2n_z+1)\epsilon/\Omega_z\approx0.01$, where $n_z$ 
is the vibrational quantum number for axial atomic oscillation, $\Omega_z$ the oscillation frequency and $\epsilon$ is the recoil 
frequency. Elastic scattering events involving no change in vibrational level are favored. The spectrum is thus Doppler-free, and we do 
not expect recoil shifts.

The complex scattering amplitude $r=|r|e^{i\phi}$ represents the global response of the atomic cloud to an incident laser 
$E_i=E_{i0}e^{i\omega_i t}$. Thus the Bragg scattered light is given by $E_s=r E_i=|r|E_{i0}e^{i\omega_i t+i\phi}$. While the amplitude 
$|r|$ is obtained via simple absorption spectroscopy, acquiring phase informations $\phi(t)$ needs heterodyning. Therefore we beat 
the Bragg scattered light with a reference beam $E_r=E_{r0}e^{i\omega_r t}$ while scanning over the resonance, $I=|E_r+E_s|^2$. We 
obtain $I\approx E_{r0}^2+|r|^2E_{i0}^2+2|r|E_{r0}E_{i0}\cos\left(\Delta\omega_i t+\phi\right)$. 
	
In order to extract amplitude and phase from a beat signal spectrum shown in Fig.~\ref{Fig3}(a), we extract the 
quadrature and the in-phase component by numerically demodulating the beat signal with $\cos{\Delta\omega_it}$ and 
$\sin{\Delta\omega_it}$. Low-pass filtering the $DC$ components in the Fourier spectrum yields $\bar{U_s}=-|r|E_{r0}E_{i0}\sin\phi$ and 
$\bar{U_c}=|r|E_{r0}E_{i0}\cos\phi$. Fig.~\ref{Fig3}(b) shows the quadrature phases. Phase and amplitude follow from 
$rE_{r0}E_{i0}(t)=(\bar{U_c}^2+\bar{U_s}^2)^{1/2}$ and $\tan\phi(t)=-\bar{U_s}/\bar{U_c}$. The result is shown in 
Fig.~\ref{Fig3}(c). We notice an absorptive profile for the reflection amplitude, which coincides with the profile 
recorded without heterodyning (shutter $S$ is closed) [see Fig.~\ref{Fig2}(c)]. The dispersively shaped phase 
profile in Fig.~\ref{Fig2}(c) exhibits a maximum phase shift on the order of $\pi$ and a distorsion due to the 
hyperfine splitting of the upper level.

To describe the observation we calculate the complex reflection coefficient $r\propto\alpha$ based on the evaluations accompanying 
Eq.~(\ref{EqBragg}) and use it to generate the beat signal shown in Fig.~\ref{Fig3}(d). This signal, if submitted to 
the same data processing as for the experimental data, yields the curves in Figs.~\ref{Fig3}(e-f). In particular, 
the amplitude and phase profile of Fig.~\ref{Fig3}(f) exactly recover the calculated complex reflection coefficient. 
To compare with the experiment, we adjust the power values in the modes $E_r\approx E_i\approx 54~$pW. We notice a good agreement, 
despite the noise appearing in the measured data. This noise is due to frequency fluctuations of the blue laser beam and to variations 
in the position of the ring cavity standing wave. 

The phase delay is intrinsically connected with the Rayleigh scattering process, which predicts a phase shift described by 
$\tan{\phi}=$Im$~\alpha/$Re$~\alpha=-\Gamma_{brg}/2\Delta_{brg}$, i.e. the phase evolves from $\phi=0$ to $-\pi$ across the resonance as 
shown in Fig.~\ref{Fig3}(f). Additional phase shifts may, in principle, result from the finite propagation time of the 
incident beam slowed down by refractive index variations in the optically dense cloud, and from multiple scattering between the atomic 
layers. In our case, the finite radial size of the scattering layers limits the effective number of layers participating in multiple 
scattering to $N_{eff}=2w_r/\lambda_{dip}\tan\beta_i\approx160$. We estimate our mean density to $n=5\times10^{11}~$cm$^{-3}$. In this 
thin grating regime, the above effects are not expected to contribute to the observed signals. We verified the validity of this 
assumption by calculating the complex reflection coefficient using the transfer matrix formalism for Bragg scattering developed by 
Deutsch \textit{et al.} \cite{Deutsch95}, and we found identical results. 

\begin{figure}[ht]
		\centerline{\scalebox{0.46}{\includegraphics{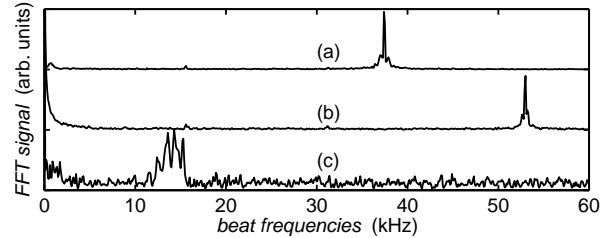}}}\caption{
		Detection of moving Bragg lattices. Shown are the Fourier spectra of 
		\textbf{(a)} the Doppler-shift $2k_{dip}v=\omega_+-\omega_-=(2\pi)37~$kHz, 
		\textbf{(b)} the reference interferometer $\Delta\omega_i=(2\pi)52~$kHz and 
		\textbf{(c)} the Bragg interferometer $\Delta\omega_s=(2\pi)15~$kHz.}
		\label{Fig4}
		\end{figure}

Our detection method is well-suited not only to probe atomic gratings, but also to detect their "motional dynamics". To demonstrate this 
we let the standing wave rotate in the ring cavity with velocity $v$ by supplying different pump frequencies, $\omega_+\neq\omega_-$. 
Fig.~\ref{Fig4} demonstrates that the Bragg scattered light is frequency-shifted by $\Delta\omega_s\equiv\omega_s-
\omega_r=\Delta\omega_i-2k_{dip}v$. The asymmetric broadening of the Fourier spectrum of the Bragg beat in Fig.~\ref{Fig4}(c) arises 
from the fact that the beat signal, being a scan over two hyperfine resonances, is non-periodic. The carrier frequency corresponds to 
the rightmost peak of the spectrum. 

In conclusion, periodic ordering in atomic clouds can have dramatic influence on the propagation and scattering of light. For thin 
atomic lattices, the Rayleigh scattered light destructively interferes in all but one direction. The resonant enhancement of Rayleigh 
scattering in this direction provided us with enough intensity to realize a "Bragg interferometer" in an atomic gas. This method may 
prove sufficiently accurate for probing interesting features of Bragg scattering in the limit where multiple reflections between 
adjacent layers are frequent, such as the occurrence of photonic band-gaps for certain ranges of light detuning or incident angle 
\cite{Deutsch95}.

Long-range order does not need to be created by periodic force fields. Under certain circumstances \cite{Kruse03b} atomic ensembles 
driven by dissipative forces spontaneously arrange themselves into propagating periodic lattices. An interesting application off our 
method of detecting moving lattices could be the study of the bunching of such systems.

\bigskip

We acknowledge financial support from the Landesstiftung Baden-W\"urttemberg.

\end{document}